# Virtual Reality Photo-based Tours for Teaching Filipino Vocabulary in an Online Class in Japan: Transitioning into the New Normal

**Roberto B. Figueroa Jr.**
robertojr.figueroa@up.edu.ph
**University of the Philippines Open University, Philippines**

**Florinda Amparo Palma Gil**
floripg@tufs.ac.jp
**Tokyo University of Foreign Studies, Japan**

**Hiroshi Taniguchi**
htaniguchi@up.edu.ph
**University of the Philippines Open University, Philippines**

**Joshze Rica Esguerra**
jlesguerra2@up.edu.ph
**University of the Philippines Open University, Philippines**

*Abstract*: When educational institutions worldwide scrambled for ways to continue their classes during lockdowns caused by the COVID-19 pandemic, the use of information and communication technology (ICT) for remote teaching has become widely considered to be a potential solution. As universities raced to implement emergency remote teaching (ERT) strategies in Japan, some have explored innovative interventions other than webinar platforms and learning management systems to bridge the gap caused by restricted mobility among teachers and learners. One such innovation is virtual reality (VR). VR has been changing the landscape of higher education because of its ability to "teleport" learners to various places by simulating real-world environments in the virtual world. Some teachers, including the authors of this paper, explored integrating VR into their activities to address issues caused by geographical limitations brought about by the heightened restrictions in 2020. Results were largely encouraging. However, rules started relaxing in the succeeding years as more people got vaccinated. Thus, some fully online classes in Japan shifted to blended learning as they moved toward fully returning to in-person classes prompting educators to modify how they implemented their VR-based interventions. This paper describes how a class of university students in Japan who were taking a Filipino language course experienced a VR-based intervention in blended mode, which was originally prototyped during the peak of the ERT era. Moreover, adjustments and comparisons regarding methodological idiosyncrasies and findings between the fully online iteration and the recently implemented blended one are reported in detail.

## INTRODUCTION

**Background of the Study**

During lockdowns caused by the COVID-19 pandemic, universities raced to implement emergency remote teaching (ERT) strategies in Japan. Some have explored innovative interventions other than webinar platforms and learning management systems to bridge the gap caused by restricted mobility among teachers and learners. One such innovation is virtual reality (VR). VR has been changing the landscape of higher education because of its ability to "teleport" learners to various places by simulating real-world environments in the virtual world. To fill in the gap brought by geographical limitations due to heightened restrictions in 2020, educators at Tokyo University of Foreign Studies (TUFS) explored integrating VR in teaching the Filipino Language to first year Japanese students (Figueroa et al., 2022).

The Filipino language was first taught in Japan at the Osaka University of Foreign Studies, now Osaka University in 1983 followed by TUFS in 1992 (Laranjo, 2020). These universities offer an entire major course in the Filipino language and Philippine-related courses. Before the pandemic, classes were held using traditional in-person classroom-based or blended pedagogy using a learning management system (LMS). Students were encouraged to join short-term language classes abroad during the long spring and summer vacations or to join one-year student exchange programs with affiliated universities abroad. These programs not only provided a more immersive experience for the learners as they used the language and interacted with the native speakers of the language they were studying, but they also increased their motivation to apply and experience first-hand what they learned inside the classroom.

Therefore, when the short-term visits and student exchange programs were canceled due to the stricter rules at the height of the pandemic in 2020, a photo-based VR tour lessons on Filipino vocabulary at TUFS was created to provide students with an immersive way of learning Filipino language and experience the Filipino culture at the comfort of their homes while being unable to physically visit the Philippines (Figueroa et al., 2022). However, rules started relaxing in 2021 and 2022 when vaccines were introduced. Thus, some fully online classes in Japan shifted to blended learning. The same happened at TUFS. With favorable feedback from students in 2020, the photo-based VR tour lessons on Filipino vocabulary were consequently integrated even in the blended offering of the course in 2021 and 2022.

**Research Questions**

With the new changes, the procedure on how the photo-based VR tour lessons were incorporated into the Filipino Language course at TUFS was revised to fit the course's evolving context. This paper aims to compare experience and related outcomes between the fully online classes in 2020 and the blended-learning implementation in 2022 by answering the following research questions.

1. How different were the satisfaction, presence, and interest felt and experienced by learners between groups who used VR tours and those who did not in each tour in 2022?
2. How different were the satisfaction and presence felt by learners who used the VR tour-based lessons between 2020 and 2022?

## RESEARCH DESIGN & METHODS

**Duration and Nature of the Study**

This longitudinal study compared 2020 and 2022 implementations of VR tour lessons. The lessons in both 2020 and 2022 spanned two months. Described as a cognitive innovation, the 2020 pilot of the VR tour lessons followed the design-based research approach where iterative design and implementation cycles were adjusted and modified based on the data collected and analyzed from each cycle.

**Context Comparison**

The two implementations had slightly different contexts. Table 1 shows slight nuances and similarities between the 2020 and 2022 implementations including the profile of students and how the classes were conducted. The participants were Japanese university students who were enrolled in the Philippine Studies Program.

There were 15 student participants in the 2020 implementation and 12 participants in the 2022 implementation. In 2020, all the photo-based VR tour lessons were held online, while the 2022 classes were both held online and during face-to-face classes. In the same year, students were divided into three groups - high immersion group (used VR goggles), moderate immersion group (did not use VR goggles but used the VR tours) and low immersion group (did not use VR goggles and VR tours; only used photo-based PowerPoint tours). In the 2022 implementation, the students were only divided into two groups, but both groups were able to experience the photo-based VR tours while using VR Goggles and the photo-based tours presented in PowerPoint presentations.

Table 1

*Contextual Data of the 2020 and 2022 Implementations*

| Variable | 2020 Implementation | 2021 Implementation |
| --- | --- | --- |
| Number of Students | 15 | 12 |
| Year Level | First Year | First Year |
| Mode | Fully Online (Synchronous) | Blended (Alternating Online and In-Person) |
| Activity Groupings | 3 (Immersive Tour, Non Immersive Tour, PowerPoint) | 2 (Immersive Tour, PowerPoint) |
| Group Composition | Group 1: 5 students<br>Group 2: 5 students<br>Group 3: 5 students | Group 1: 6 students<br>Group 2: 6 students |

**Sequence of Activities**

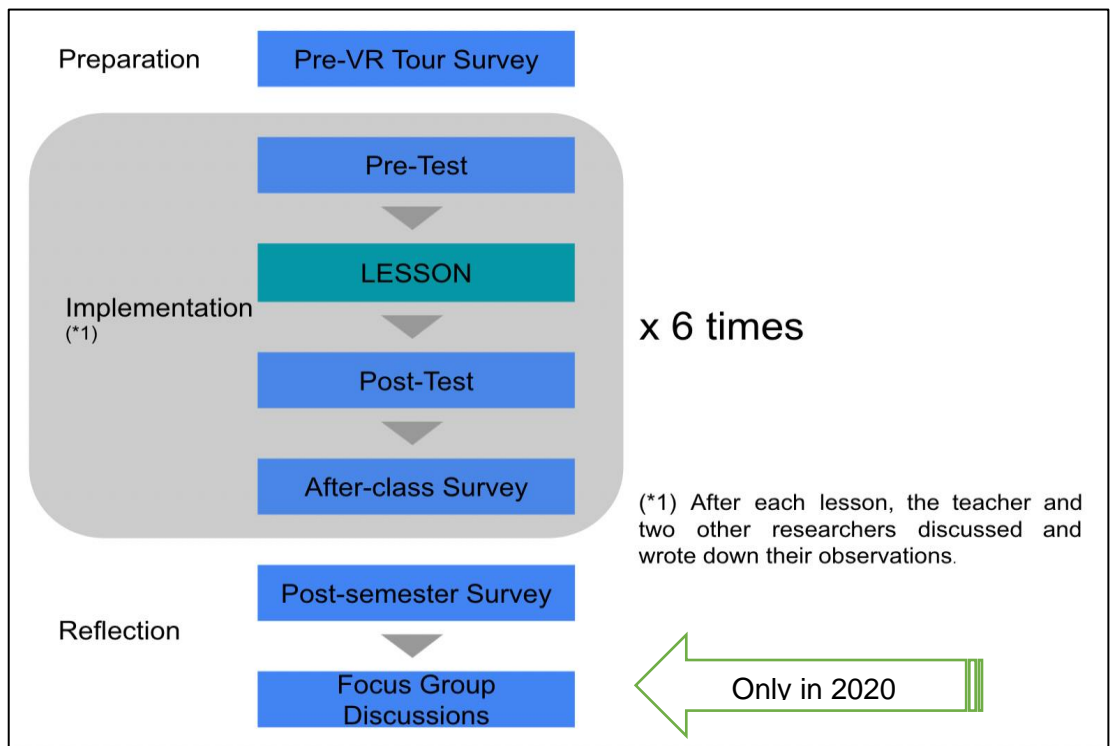

*Figure 1.* Procedural Diagram of the 2020 and 2022 Implementations as Illustrated in Figueroa et al. (2022)

The sequence of activities were the same in both the 2020 and 2021 implementations as shown in Fig. 1. The procedural diagram was directly lifted from Figueroa et al. (2022). As illustrated, a survey was given to students at the beginning of the semester before they could experience the VR or PowerPoint presentation tours. The steps in the darker square represent activities that are conducted in class. There were six classes conducted in both implementations, which included a pre-test, the lesson proper that involved the tours, a post-test, and an after class survey. At the end of the semester, students were asked to reflect on the whole experience through a post-semester survey and focus group discussions. The only difference during the 2022 implementation was that there was no more focus group discussion conducted.

**Group Configuration**

Another major difference between the two implementations is the grouping configuration. Three groups were formed in 2020 (high, medium, and low). The high immersion group consisted of students who experienced VR tours using their smart phones with VR goggles delivered to their homes. The medium immersion group consisted of students who experienced VR tours without the VR goggles. The low immersion group consisted of students who experienced PowerPoint-based tours with the same content as the VR tours. The grouping was only changed once, after the first lesson where some students reported their smartphones not working with the goggles. However, in the five succeeding lessons, the groupings and their assigned activities did not change (Figueroa et al., 2022). In contrast, the implementation in 2022 only involved two groups. As illustrated in Fig. 2, in the first three lessons, Group 1 experienced VR tours with goggles (VR Group) while Group 2 experienced PowerPoint-based tours (Non-VR Group). In the second three lessons, Group 2 became the VR group and Group 1 became the Non-VR Group. This was done so that all the students may be able to experience both types of activities.

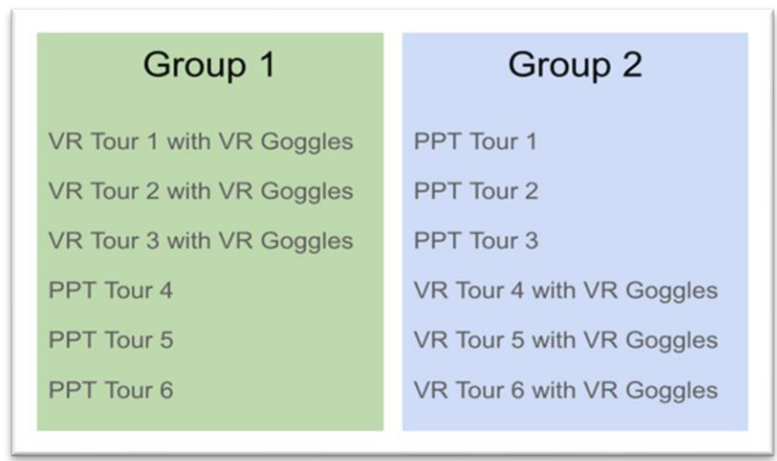

*Figure 2.* Group Configuration in 2022 Implementation

**Platform Selection for Immersive Open Pedagogical Activities**

In this section, we shall describe the platforms used in the two iterations of the study. Kuula is a web-based software that makes it easy to create 360° virtual tours. The free basic plan allows level correction and retouching of images, while paid plans ranging from 16 to 48 US Dollars per month include audio support, unlimited uploads, unlisted and password-protected tours, custom icons and fonts, and analytics (Kuula, n.d.).

A free alternative to Kuula with audio support is StorySpheres, a website created by Grumpy Sailor with the help of Google's Creative Lab in 2014 (Story Spheres, n.d). A user must upload 1 JPG/JPEG image and at least 1 MP3 audio file, with the total size of all files below 15 MB. In addition to having a background sound, audio hotspots can easily be added and positioned using the slider, as shown in Fig. 3.

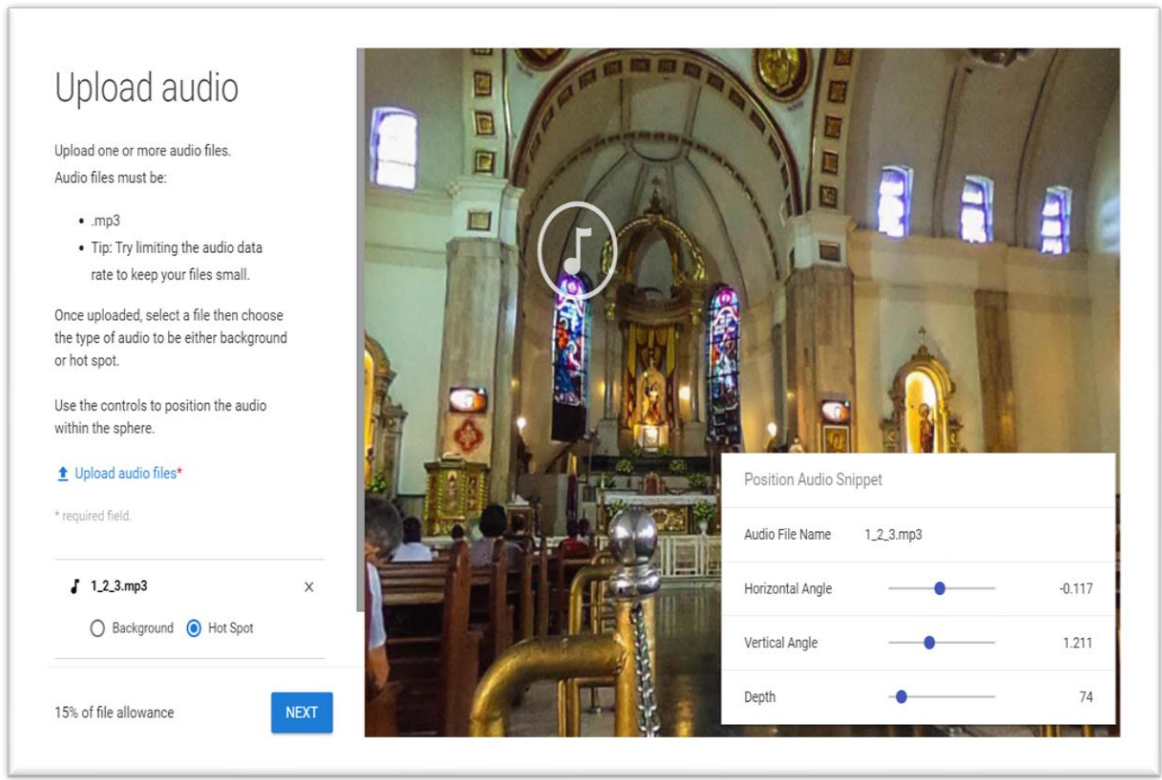

*Figure 3.* Using and Positioning Hotspots to Play Audio Narrations in Story Spheres

For those with HTML and JavaScript knowledge, A-Frame (https://aframe.io/) is a notable option for more freedom in developing 360° tours. It is a web framework based on top of HTML for building VR experiences with only text-editing software and a web browser needed. When developing a virtual tour, JavaScript can be used to change the image, music, and hotspot locations upon the click of a user. Since it requires coding, it will allow for more freedom and customization in the tours. For example, all paid features in Kuula can be done in A-Frame, with the only limitation being the learning curve. A finished A-Frame project can be deployed to a user's server for personal or company branding, or online Integrated Development Environments (IDEs) with hosting such as Glitch (https://glitch.com/). Table 2 shows a comparative summary of the main features of the three platforms presented in this section.

**Converting from Kuula to A-Frame**
Kuula was a viable option in the 2020 implementation because of its capability to facilitate rapid prototyping. However, because of the recurring costs of maintaining a paid account, A-Frame was chosen to migrate the developed VR tours for sustainability and was eventually used in the 2022 implementation.

The first step was to retrieve the 360° images from Kuula by clicking the Download link at the bottom of the Edit pane and then saving the image. Recognizable faces on all photos were blurred using Adobe Fresco. The narrations had to be recorded using Audacity since the Kuula platform did not allow audio files to be downloaded from its tours.

The index page with portals used a 360° panoramic image as the initial source of the <a-sky> element. There were multiple portals, each one an <a-circle> element with its source and the image representing the destination. Behind it is a white <a-circle> to mimic an outline. Since A-Frame does not have support for non-alphanumeric text, Japanese characters were added by importing a Multi-channel Signed Distance Font (MSDF) file that was generated online.

Table 2

*Comparison of the VR Tour Platforms*

| Platform | Price | Features | Limitations |
|---|---|---|---|
| Kuula | Free | • Retouch images<br>• Level correction<br>• Private tours<br>• Choose transition type<br>• Add images and hotspots<br>• Hotspots can open video/text cards and URLs | • No audio support<br>• Max 100 uploads per month<br>• Max 25 images per batch upload |
|  | 16-20 USD per month | • Allows audio files<br>• Walkthrough mode<br>• Unlisted tours<br>• Custom icons and fonts | • 360° videos not supported |
|  | 36-48 USD per month | • Custom domain<br>• Password-protected tours<br>• Analytics |  |
| StorySpheres | Free | • Allows audio files in the background or hotspot | • Stitching is not seamless<br>• Up to 15MB total file size<br>• Requires at least 1 audio file<br>• Video files are not supported<br>• Cannot add text |
| A-Frame | Free | • Allows for more freedom and customization<br>• 360° videos supported<br>• Can host on own or cloud servers | • Requires coding<br>• Learning curve |

When a user hovers on a portal, there will be a preview (see Fig. 4) by displaying the name of the place and temporarily changing the <a-sky> source with the use of JavaScript. The animation component was utilized to make the transition smoother. Clicking a portal will redirect the browser to the tour of that location. Fig. 5 shows the interface of the tour when entering VR mode on a mobile browser.

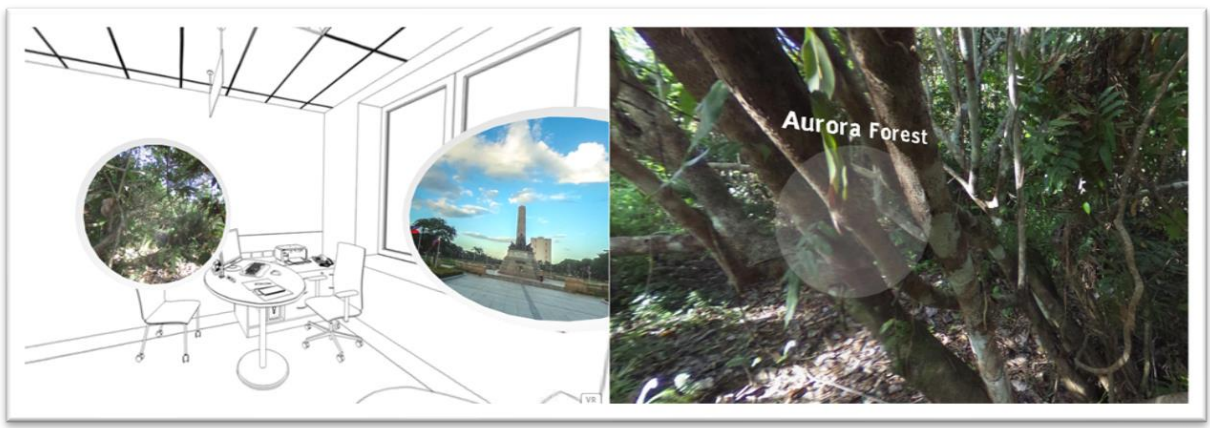

*Figure 4.* The User Interface Before and After Hovering on a Portal

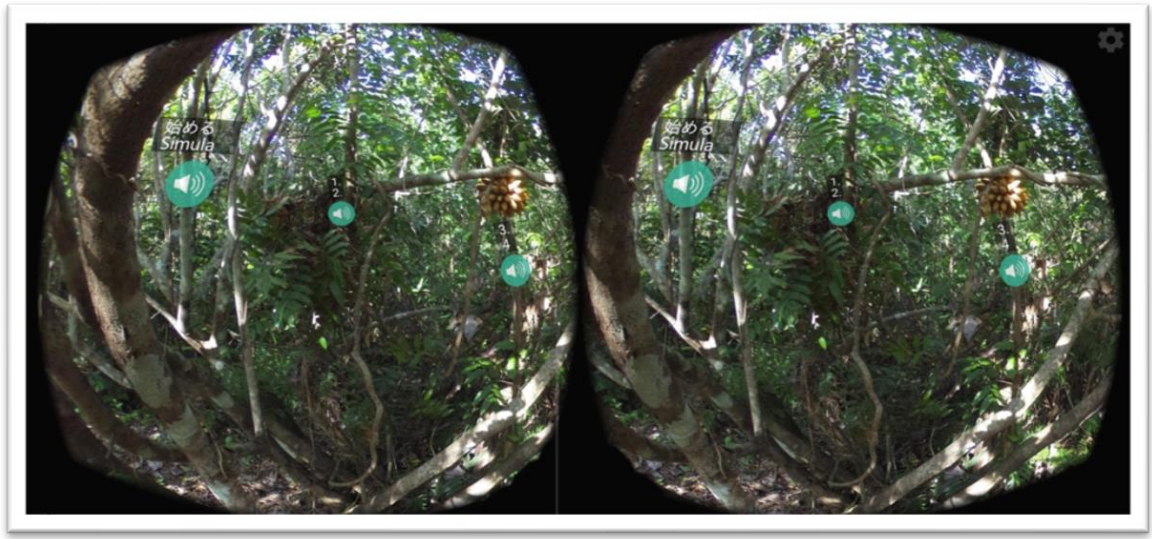

*Figure 5.* Viewing the Tour on a VR-Ready Mobile Phone

Each tour includes multiple narrations that will play when its corresponding audio button is clicked. Audio buttons are <a-image> hotspots that are mapped with the help of the A-Frame Inspector (see Fig. 6), by dragging it to the corresponding position and copying the coordinates to the position attribute in the code. The look-at component is used to easily change the angle so that it will always face the user. When the button is clicked, the script will change the sound attribute of the a-sky to the narration and toggle the player.

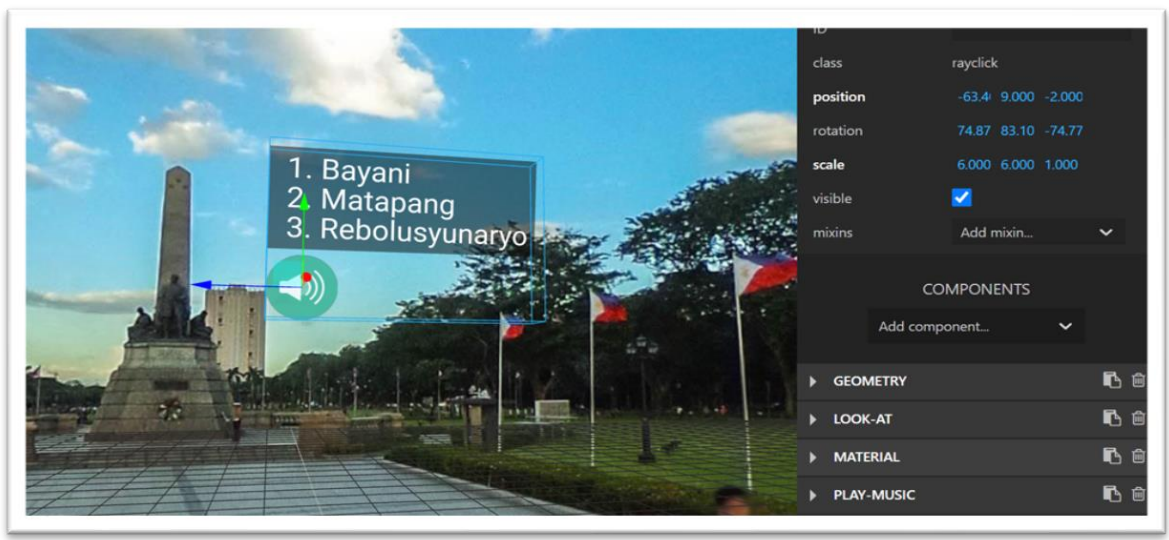

*Figure 6.* Getting the Position Coordinates in A-Frame Inspector

While the created tours are on separate web pages for easier sharing and access, another approach would be to use a single webpage to host all tours. This can be done by using JavaScript to change the source of the <a-sky> tag and the coordinates and identifiers of each audio hotspot with each click on the portal. However, since the tours were non-contiguous and were presented separately, they were developed as separate pages.

**Data Collection**

The data used in this study include the results of six after-class surveys in the (1) 2020 implementation and the (2) data collected during the photo-based VR tour lessons held in the first semester spanning from May to June in 2022. The results of the pre-test and post-test quizzes were not included as they were not included in the scope of the study. All questionnaires contain both Likert-type items and open-ended questions. Data (1) was analyzed to answer RQ1 while data (1) and (2) were compared to answer RQ2. Fig. 7 was a table lifted from an appendix of the previous publication (Figueroa et al., 2022), which lists the after-class survey items used in both the 2020 and 2022 implementations. Among these, only items two, four, and 12 were used for this study. Item two, which was boxed in red in the figure, represented satisfaction. Item four, which was boxed in blue, represented interest and item 12, which was boxed in green, represented presence. All the items were translated in the Japanese language. Face validity and language expert consultation were conducted for the three items. While there was no other validity and reliability tests conducted for the interest and satisfaction items, the presence item was a slightly modified version of the single-item measure proposed and validated by Bouchard et al. (2004).

*Figure 7.* After-class Survey Questions in the 2020 and 2022 Implementations

**Data Analysis**

To answer the first research question, summary statistics were generated for satisfaction, presence, and interest among students of the two groups in each of the six lessons to see whether there are trends regarding differences. Statistical significance was determined by performing the Mann-Whitney U test in each lesson using the *stats* library in R (R Core Team, 2012). To answer the second question, summary statistics and boxplots were generated for satisfaction, presence, and interest among students of lessons two through five in 2020 and 2022. Statistical significance per lesson was determined by performing the Mann-Whitney U.

# RESULTS

**RQ 1: How different were the satisfaction, presence, and interest felt and experienced by learners between groups who used VR tours and those who did not in each tour in 2022?**

**Lesson 1**
Table 3 compares the medians of the satisfaction, presence, and interest ratings of students in the VR and non-VR groups in lesson 1.

Table 3

*Comparison of Medians of Student Ratings between 2 Groups in Lesson 1*

| Group | Satisfaction | Presence | Interest |
|---|---|---|---|
| VR (1) | 10 | 9.5 | 10 |
| Non-VR (2) | 8 | 6 | 7.5 |

The Mann Whitney U test indicated that satisfaction ratings were greater for students in the VR group ($Mdn = 10$) than those in the non-VR group ($Mdn = 8$), $U = 35, p = .006$. It also indicated that presence ratings were greater for students in the VR group ($Mdn = 9.5$) than those in the non-VR group ($Mdn = 6$), $U = 36, p = .004$. However, interest ratings were not statistically significantly different between the two groups.

**Lesson 2**
Table 4 compares the medians of the satisfaction, presence, and interest ratings of students in the VR and non-VR groups in lesson 2.

Table 4

*Comparison of Medians of Student Ratings between 2 Groups in Lesson 2*

| Group | Satisfaction | Presence | Interest |
|---|---|---|---|
| VR (1) | 10 | 9.5 | 10 |
| Non-VR (2) | 8 | 6 | 7.5 |

The Mann Whitney U test indicated that none of the three variables were statistically different between the two groups.

**Lesson 3**
Table 5 compares the medians of the satisfaction, presence, and interest ratings of students in the VR and non-VR groups in lesson 3.

Table 5

*Comparison of Medians of Student Ratings between 2 Groups in Lesson 3*

| Group | Satisfaction | Presence | Interest |
|---|---|---|---|
| VR (1) | 10 | 10 | 10 |
| Non-VR (2) | 8 | 7 | 8 |

The Mann Whitney U test indicated that presence ratings were greater for students in the VR group ($Mdn = 10$) than those in the non-VR group ($Mdn = 7$), $U = 31, p = .018$. However, satisfaction and interest ratings were not statistically significantly different between the two groups.

**Lesson 4**

Table 6 compares the medians of the satisfaction, presence, and interest ratings of students in the VR and non-VR groups in lesson 4.

Table 6

*Comparison of Medians of Student Ratings between 2 Groups in Lesson 4*

| Group | Satisfaction | Presence | Interest |
|---|---|---|---|
| VR (2) | 9.5 | 9.5 | 9 |
| Non-VR (1) | 10 | 10 | 10 |

The Mann Whitney U test indicated that none of the three variables were statistically different between the two groups.

**Lesson 5**

Table 7 compares the medians of the satisfaction, presence, and interest ratings of students in the VR and non-VR groups in lesson 5.

Table 7

*Comparison of Medians of Student Ratings between 2 Groups in Lesson 5*

| Group | Satisfaction | Presence | Interest |
|---|---|---|---|
| VR (2) | 10 | 10 | 10 |
| Non-VR (1) | 10 | 10 | 10 |

The Mann Whitney U test indicated that none of the three variables were statistically different between the two groups.

**Lesson 6**

Table 8 compares the medians of the satisfaction, presence, and interest ratings of students in the VR and non-VR groups in lesson 6.

Table 8

*Comparison of Medians of Student Ratings between 2 Groups in Lesson 6*

| Group | Satisfaction | Presence | Interest |
|---|---|---|---|
| VR (2) | 8.5 | 8.5 | 8 |
| Non-VR (1) | 10 | 10 | 10 |

`The Mann Whitney U test indicated that none of the three variables were statistically different between the two groups.

**RQ 2: How different were the learning outcomes and attitudes of learners who used the VR tour-based lessons between 2020 and 2022?**

Fig. 8 compares 2020 and 2022 boxplots of satisfaction, presence, and interest ratings that were aggregated across lessons two to six. It could be seen that the ratings of satisfaction, presence, and interest in 2022 were generally higher than the ratings of the three variables in 2020.

The Mann Whitney U test conducted per lesson confirmed this trend in the second and third lessons. In the second lesson, there was a statistically significant difference in satisfaction ratings between 2020 ($Mdn = 9$) and 2022 ($Mdn = 10$), $U = 7.5$, $p = .02$. In the same lesson, there is a statistically significant difference in presence ratings between 2020 ($Mdn = 8$) and 2022 ($Mdn = 10$), $U = 9.5$, $p = .05$.

In the third lesson, there was a statistically significant difference in satisfaction ratings between 2020 (*Mdn* = 8) and 2022 (*Mdn* = 10), $U = 7.5$, $p = .02$. In the same lesson, there is a statistically significant difference in presence ratings between 2020 (*Mdn* = 8) and 2022 (*Mdn* = 10), $U = 9.5$, $p = .003$. None of the other lessons had statistically significant differences in presence and satisfaction ratings. Furthermore, there were no statistically significant differences in interest ratings between the two-year offerings.

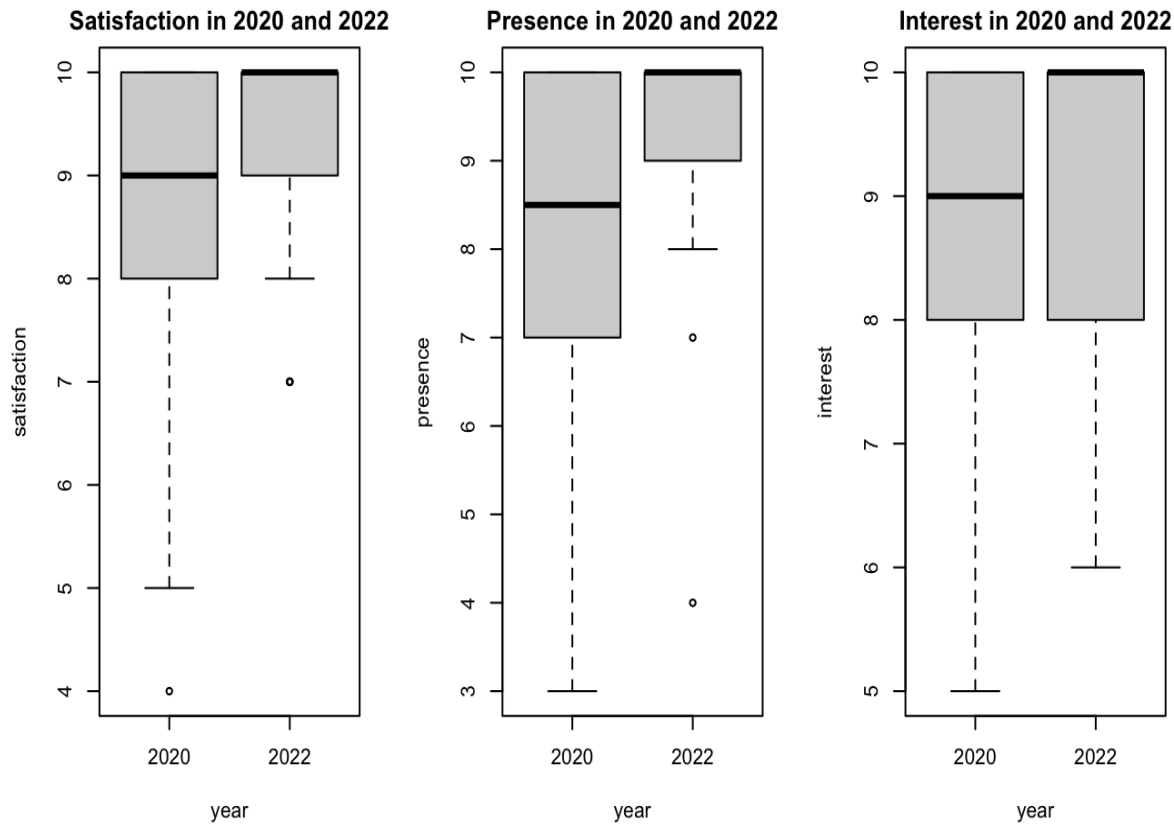

*Figure 8.* Comparative Boxplots of Aggregated Ratings of Satisfaction, Presence, and Interest in 2020 and 2022

## DISCUSSION

The findings revealed very enlightening trends in similarities and differences between the activities implemented in 2020 and 2022.

**The Novelty of VR Tours**

The statistically significant difference in presence, interest, and satisfaction between VR and Non-VR Groups in the first lesson of the 2022 implementation showed that the VR tours piqued the students' interest, provided more spatial presence, and gave them a better experience than in the PowerPoint-based tours. However, this was not evident in the succeeding lessons. This may be explained by novelty, which was found to increase the interest among participants and viewed by motivational researchers as one of its dimensions or components (Deci, 1992; Sun et al., 2008). However, novelty wanes through time (Spielberger & Starr, 1994). This may have happened in the succeeding lessons. Unlike in the 2020 implementation where data supported interest in the succeeding lessons, data which could support this trend in the 2022 implementation was yet to be analyzed, thereby posing a significant limitation of this study. However, the findings of this study highlighted that a VR tour is a practical activity for gaining attention, which was recommended as an initial step in effective teaching according to Gagne's nine events of instruction (Schunk, 2012).

**In-Person Orientation Benefits**

Another revelation was that 2022 implementation of the VR-based activities in blended mode yielded higher presence and satisfaction ratings than that in the purely online mode in 2020. These findings showed the advantage of conducting VR-based activities in blended settings compared to purely remote ones. The learning curve and technical challenges in training students to use a VR device in a purely remote environment may have blunted the motivational benefits that could have been obtained from using these novel technologies. The blended nature of the classes in 2022 enabled the teacher to support students in using the VR devices in person. They could still access it during the online sessions, but they were already well acquainted with the technology through the in-person orientation. The importance of ensuring that students are comfortable in using an instructional technology has been echoed by studies in technology readiness (Hubbard, 2013; Ngampornchai & Adams, 2016; Warden et al., 2022) . Therefore, having an initial in-person session to help students get acquainted with VR devices and applications for VR-based learning activities even in purely online learning settings would be extremely helpful as the technology is still not that common.

The piloting and prototyping nature of the implementation in 2020 could also be attributed for this observation. During that time, many of the problems encountered by students were still unknown and had to be discovered. Those problems have already been addressed in the 2022 implementation. This confirms the practical benefits of employing a design-based research approach in 2020, which was characterized by iterative cycles of design, enactment, analysis, and redesign in a single setting over a period (Design-Based Research Collective [DBRC], 2003).

## CONCLUSION

With many of the traditional universities embracing blended learning after implementing fully online classes during the height of the COVID-19 pandemic, opportunities for improving students' experience in technology-enhanced learning can be explored. In this paper, the findings from a study involving a method of learning a foreign language in a remote teaching context through VR tours in 2020 and changes in the 2022 implementation were presented. While limitations persist regarding generalizability and the need for qualitative data that could support earlier findings, the study may provide practical insights regarding the advantage of in-person technical training and the benefits of piloting a method using the design-based research approach.